\UseRawInputEncoding 
\documentclass{INTERSPEECH2023}
\usepackage{subfigure,cite,color, hyperref}

 \interspeechcameraready

\title{BASEN: Time-Domain Brain-Assisted Speech Enhancement Network with Convolutional Cross Attention in Multi-talker Conditions}
\name{Jie Zhang$^1$, Qing-Tian Xu$^2$, Qiu-Shi Zhu$^1$, Zhen-Hua Ling$^1$}
\address{$^1$NERC-SLIP, University of Science and Technology of China (USTC)\\
 $^2$School of Electronics and Information Engineering, Sichuan University}
\email{\{jzhang6; zhling\}@ustc.edu.cn, xuqingtian@stu.scu.edu.cn, qszhu@mail.ustc.edu.cn}

\begin{document}
\setlength{\abovedisplayskip}{3.2pt}
\setlength{\belowdisplayskip}{3.2pt}
\maketitle
 
\begin{abstract}
 Time-domain single-channel speech enhancement (SE) still remains challenging to extract the target speaker without any prior information on multi-talker conditions. It has been shown via auditory attention decoding that the brain activity of the listener contains the auditory information of the attended speaker. In this paper, we thus propose a novel time-domain brain-assisted SE network (BASEN) incorporating electroencephalography (EEG) signals recorded from the listener for extracting the target speaker from monaural speech mixtures. The proposed BASEN is based on the fully-convolutional time-domain audio separation network. In order to fully leverage the complementary information contained in the EEG signals, we further propose a convolutional multi-layer cross attention module to fuse the dual-branch features. Experimental results on a public dataset show that the proposed model outperforms the state-of-the-art method in several evaluation metrics. The reproducible code is available at \href{https://github.com/jzhangU/Basen.git}{{\color{blue}https://github.com/jzhangU/Basen.git}}.
 \end{abstract}

\section{Introduction}
It is natural for human auditory systems  to extract the auditory information of the attended speaker while attenuate competing speakers in multi-talker conditions. This facilitates various  applications in e.g., speech recognition~\cite{Jinyu-et-al:scheme}, hearing aids~\cite{zhao2018deep}, speech synthesis~\cite{krishna_advancing_2020}, etc. However, speech enhancement (SE) suffers from this efficacy in such multi-talker scenarios, which is often-used  to improve the speech quality or intelligibility in speech interaction systems. 
The focus of this work is on the monaural SE issue, as the single-microphone setup is more flexible and economic than microphone arrays~\cite{benesty2008microphone} in configuration. 
Conventional well-established monaural SE methods mainly depend on statistical modeling~\cite{loizou2007speech}, which usually can perform well in stationary noise conditions. But the performance drops rapidly in case of non-stationary noises (e.g., cocktail parties), as it becomes difficult to track the noise statistics.

Recently, deep neural networks (DNNs) have been successfully applied in many fields including SE, which can achieve an even better listening performance than conventional counterparts, particularly in non-stationary cases.  The DNN-based methods can be generally categorized into  time-domain and time-frequency (T-F) domain designs.  The former is mainly based on the observation that in the short-time Fourier transform (STFT) domain the speech and noise patterns are more distinguishable, such that the training target can be learned, which is then multiplied with the noisy speech representation to recover the speech magnitude.  The training target can be masking-based, e.g., ideal ratio mask~\cite{wang2014training}, phase sensitive mask~\cite{erdogan2015phase}, and mapping-based, e.g., spectral magnitude and log-power spectrum~\cite{xu2014regression} corresponding to the spectral representation of clean speech. However, T-F domain methods only enhance the spectral magnitude and keep the phase unchanged. In order to incorporate the phase process complex spectral mapping~\cite{tan_learning_2020} and complex ratio mask~\cite{williamson_time-frequency_2017} were thus proposed, where  the magnitude and phase can be recovered  implicitly, but the magnitude distortion might contradict the phase prediction~\cite{wang2021compensation,li2022taylor}.

The time-domain methods can avoid processing the complex spectral and directly predict the clean speech waveform from the noisy input mixtures~\cite{pandey_dense_2021}. The magnitude and phase are thus jointly optimized. For example, the fully-convolutional time-domain audio separation network (Conv-TasNet) was proposed in~\cite{luo_conv-tasnet_2019} (i.e., an extension of TasNet~\cite{luo2018tasnet}), which uses a linear encoder to generate a speech representation optimized for separating individual speakers, applies a set of weighting functions (masks) to the encoder output and finally utilizes a linear decoder to recover the speech waveform.  In~\cite{pandey2019new}, a convolutional neural network (CNN) based time-domain SE framework was built, while the loss functions are calculated in the STFT domain to avoid the distortion  of the phonetic information in the estimated speech. It was shown that usually the time-domain methods have a smaller model size and a much shorter latency, which would be more appropriate for real-time applications.

\begin{figure*}[h!]
  \centering
    \includegraphics[width=0.8\textwidth]{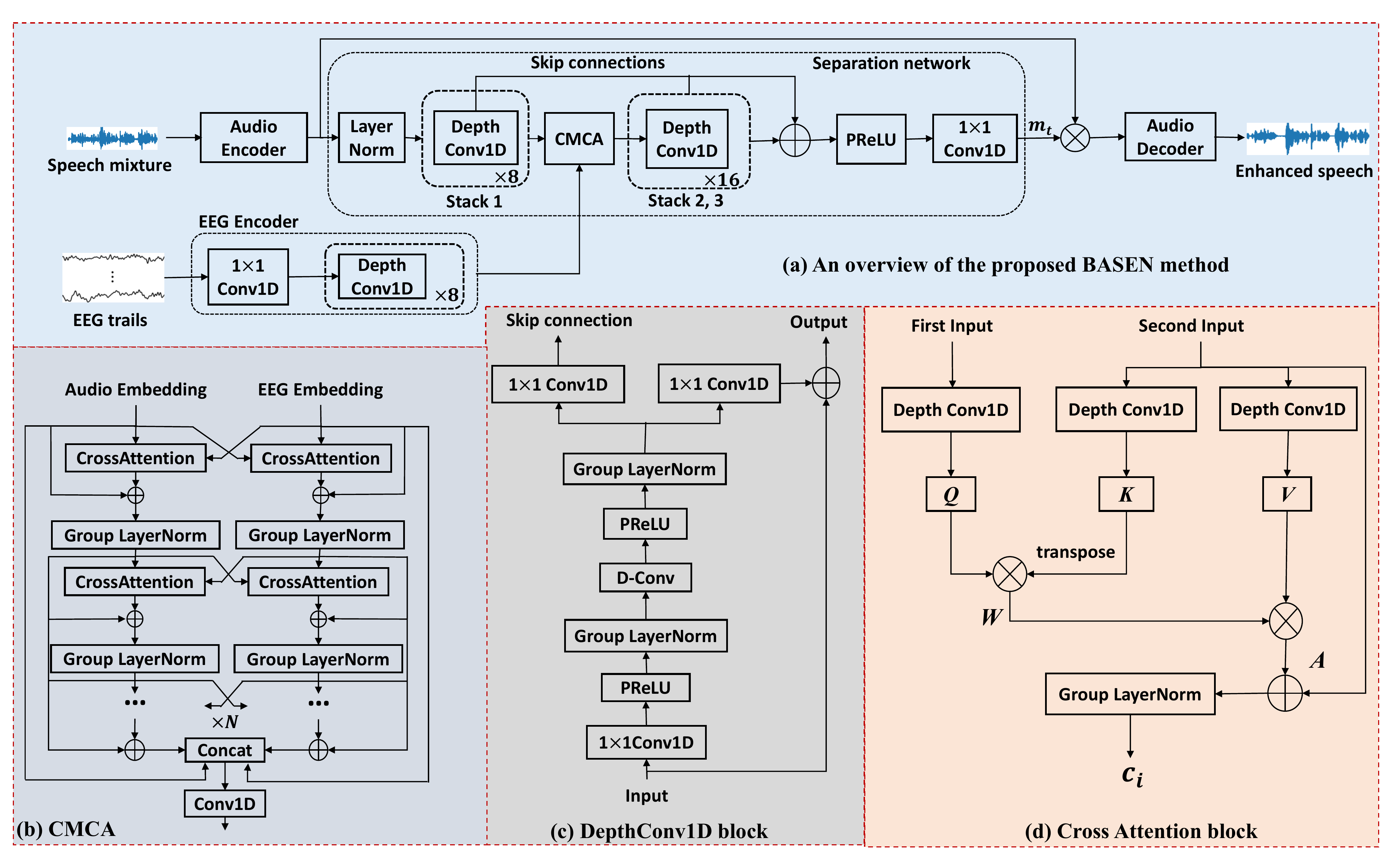}
    \vspace{-0.2cm}
  \caption{The  proposed BASEN model: (a) an overview,  (b) CMCA, (c) DepthConv1D block, and (d) cross attention block.}
  \label{fig:model}
      \vspace{-0.4cm}
\end{figure*}
However, these approaches heavily depend on prior information on, e.g., the target speaker and/or the number of speakers~\cite{hu2021speech}. The remaining challenge is that {\it how to estimate the target speech signal in multi-talker conditions without these prior information?}
It was  shown in \cite{mesgarani_selective_2012,osullivan_attentional_2015,aroudi_auditory_2016,osullivan_neural_2017,aroudi2021closed,geirnaert_electroencephalography-based_2021,han2019speaker} that in multi-talker conditions, the attended speaker can be decoded from the brain activity of the listener, which is known as auditory attention decoding (AAD). This means that speech separation has to be performed in prior to AAD. It was shown that the AAD accuracy strongly depends on the length of the decision window, which decreases drastically with a shorter decision window. Also, this cascaded scheme requires a large computational cost. 

As the speech of the attended speaker is the estimation goal, it is more natural to directly use the brain activity as an extra modality similarly to multi-model speech separation~\cite{wang2018voicefilter,ochiai2019multimodal,li2020listen,michelsanti2021overview}. Three brain-based SE methods were thus proposed, e.g., brain-informed speech separation (BISS)~\cite{ceolini_brain-informed_2020}, brain-enhanced speech denoiser (BESD) and U-shaped BESD (UBESD) \cite{hosseini_end--end_2022}.
For BISS, the estimated speech envelope is first estimated from electroencephalography (EEG) signals and then fused with the speech mixture in a target extraction network to recover the attended speech signal. The BESD and UBESD methods follow a dual-branch end-to-end scheme, which was shown to be more effective than BISS. However, in these methods the binary modalities are mainly fused using concatenation, and the cross-modal complementary information is insufficiently leveraged. 
In this paper, we therefore propose an end-to-end time-domain brain-assisted SE network (BASEN), which is built based on the typical Conv-TasNet backbone~\cite{luo_conv-tasnet_2019}. First, we use two encoders to extract the speech and EEG embeddings, respectively from the speech mixture and EEG trials. In order to explore the complementary information (e.g., the attended speaker), we design a convolutional multi-layer cross attention (CMCA) module to deeply fuse the embeddings, which are then fed into the decoder to recover the final speech waveform. Experimental results on a public dataset show the superiority of the proposed method over the current state-of-the-art UBESD model.

    \vspace{-0.1cm}
\section{Methodology}
    \vspace{-0.1cm}
{\it \textbf{Problem formulation:}}
The proposed BASEN model is graphically shown in Figure \ref{fig:model}(a), which is based on the popular Conv-TasNet \cite{luo_conv-tasnet_2019}. The original Conv-TasNet basically follows the end-to-end framework, consisting of an audio encoder, a separation network and a decoder. The BASEN additionally incorporates an EEG branch and uses the CMCA module for deep dual-modal feature fusion.
Given the noisy speech signal $x$, the audio encoder first transforms it into an embedding sequence  $w_{x}$. The EEG branch exploits an EEG encoder to extract the EEG embedding from the recorded EEG trials. That is, 
\begin{equation}
    w_{x}={\rm AudioEncoder}(x), \quad
    e_{x}={\rm EEGencoder}(e_{t}).
\end{equation}
Both audio and EEG embeddings are input to the separation network to estimate the source-specific masks, given by
\begin{equation}
    [m_{0},\cdots,m_{T-1}]={\rm Separator}(w_{x},e_x),
\end{equation}
where $T$ denotes the number of existing sources. For SE, $T$ = 2, i.e., the target speech and non-target additive noise, while for the general speech separation task $T$ might be greater than 2. Note that the audio and EEG features have to be fused in the separator using the proposed CMCA module. Finally, the reconstructed  speech signal is obtained by mapping the audio embedding using the source-specific masks, i.e.,
\begin{equation}
    \hat{s}_t={\rm Decoder}(w_{x}\odot m_t), \forall t\in\{1,\ldots,T\},
\end{equation}
where $\odot$ denotes the element-wise multiplication. 

{\it \textbf{Audio and EEG encoders:}}
The audio encoder consists of a couple of 1-dimensional (1D) convolutions for downsampling to extract the audio feature. The corresponding audio decoder exhibits a mirror structure of the audio encoder, which has the same number of 1D transposed convolutions with a stride of 8 to perform upsampling.

The EEG encoder at the lower branch of  Figure~\ref{fig:model}(a)  is composed of one 1D convolution with a stride 8 to downsample the EEG trials  and a stack of Depth-wise 1D convolutions proposed in \cite{luo_conv-tasnet_2019}, which has 8 layers and each layer has residual connections to help form multi-level features. 
The Depth-wise 1D convolution block is shown in Figure~\ref{fig:model}(c), which decouples the standard convolution operation into two consecutive operations, i.e., a Depth-wise convolution followed by a pointwise convolution. As such,  the model size can be largely reduced. Note that the Depth Conv1D module has eight basic layers and each sets a  residual connection to obtain a multi-level feature output.

{\it \textbf{Separation network:}}
The separation network aims to predict the target speaker's mask $m_{t}$ and perform feature fusion for the audio and EEG embeddings. It mainly consists of three stacks of Depth-wise 1D convolutional blocks and a CMCA-based cross attention mechanism for dual-modal feature fusion and synchronization. 
 In each DepthConv1D stack, there are $D$ layers of DepthConv with an exponential growth of the dilation factor $2^d$, where $d \in \{0, . . . , D-1\}$. 

Letting the output of the first stack be denoted by $a_{x} ={\rm Stack}_{1}(w_{x})$,
the fused feature output by the CMCA can be given by
$
    c_{x}={\rm CMCA}(a_{x},e_{x}),
$
which is then fed into two sequential DepthConv1D stacks, and the corresponding transformed feature is $
    d_{x}={\rm Stack}_{3}({\rm Stack}_{2}(c_{x})).
$
Note that this intermediate feature has to be further summed with the skip connections from the mentioned three stacks, which will be passed through a parametric rectified linear unit (PReLU) as well as a $1\times 1$ convolution to resolve the final target-specific mask $m_{t}$.

{\it \textbf{CMCA:}}
The CMCA module in Figure~\ref{fig:model}(b) is designed for feature fusion, which consists of $N$ layers of coupled cross attention blocks with skip connections and group normalization being placed between two adjacent layers. The cross attention block is depicted in Figure~\ref{fig:model}(d). The left branch of CMCA deals with the audio stream and the right branch copes with the EEG modality. In layer $i$, the inputs of two branches are denoted by $e_{i-1}$ and $a_{i-1}$, respectively. Note that in case $i=0$, $e_0=e_x$ and $a_0=a_x$. Let the right cross attention block be denoted as ${\rm CrossAtt}_r$ and the left block as  ${\rm CrossAtt}_{l}$, respectively. Similarly, we define ${\rm GroupN}_l$ and ${\rm GroupN}_r$. The outputs of layer $i,\forall i\ge 1$ can be represented as
\begin{align}
    a_{i}&={\rm GroupN}_{l}(a_{i-1}+{\rm CrossAtt}_{l}(e_{i-1},a_{i-1},a_{i-1})),\\
    e_{i}&={\rm GroupN}_{r}(e_{i-1}+{\rm CrossAtt}_{r}(a_{i-1},e_{i-1},e_{i-1})).
\end{align}
The audio-related and EEG-related layer-wise features of all layers are then added together, which will be concatenated with the original audio-EEG embeddings $(a_x,e_x)$ over the channel dimension. The concatenated result will be sent into a 1D convolution to construct the final fused dual-modal feature.

Similarly to the self-attention mechanism~\cite{vaswani2017attention}, the considered  cross attention block includes three key components: query $Q$, key $K$ and value $V$, where $\{K, Q, V\}\in \mathbb{R}^{C\times L}$ with $C$ and $L$ denoting the number of channels and the length of embeddings, respectively. We first use 3 Depth-wise 1D convolution to transform the input (i.e., $a_{i-1}$ and $e_{i-1}$) into $K$, $Q$ and $V$.  From Figure~\ref{fig:model}(d), it is clear that $K$ and $V$ come from the same sequence, while $Q$ is from the other. Specifically, the correlation between  $Q$ and $K$ is first calculated as $W=QK^T$,
where $W\in \mathbb{R}^{C\times C}$ stands for the attention weights. Next, the correlation scores are converted into probabilities using the softmax function. Then the attention weight $W$ is multiplied with the value $V$ to obtain $ A=WV$, which is then added with the residual connections and passed through a group normalization, leading to the output of the $i$-th layer of CMCA.

{\it \textbf{Loss function:}}
In this work, we use the scale-invariant signal-to-distortion ratio (SI-SDR) \cite{le2019sdr} to measure the loss function, which was shown to be a well-performed general-purpose loss function for the time-domain SE~\cite{kolbaek2020loss}, defined as
\begin{equation}
{\rm SI}\text{-}{\rm SDR}=10\log_{10}{\Vert x_{\rm target}\Vert ^{2}}/{\Vert x_{\rm res}\Vert ^{2}},
\end{equation}
where $x_{\rm target}=\frac{\hat{s}_t^Ts}{\Vert s\Vert ^2}s$ and $x_{\rm res}=x_{\rm target}-\hat{s}_t$ with $s$ and $\hat{s}_t$ standing for the target speech and the reconstructed speech, respectively. Scaling the target speaker ensures that the SI-SDR is invariant to the scale of the reconstruction, which is important for the stability of model training as the scale of the target speaker might be changed after processing. 
As in general the higher the SI-SDR, the higher the speech quality, the negative SI-SDR is thus taken as the loss function for training. In addition, we use the Adam optimizer for model training.
  \begin{figure*}[!t]
  \centering
  \subfigure[UBESD~\cite{hosseini_end--end_2022}]{
  \centering
    \includegraphics[width=0.45\textwidth,height=0.4\textwidth]{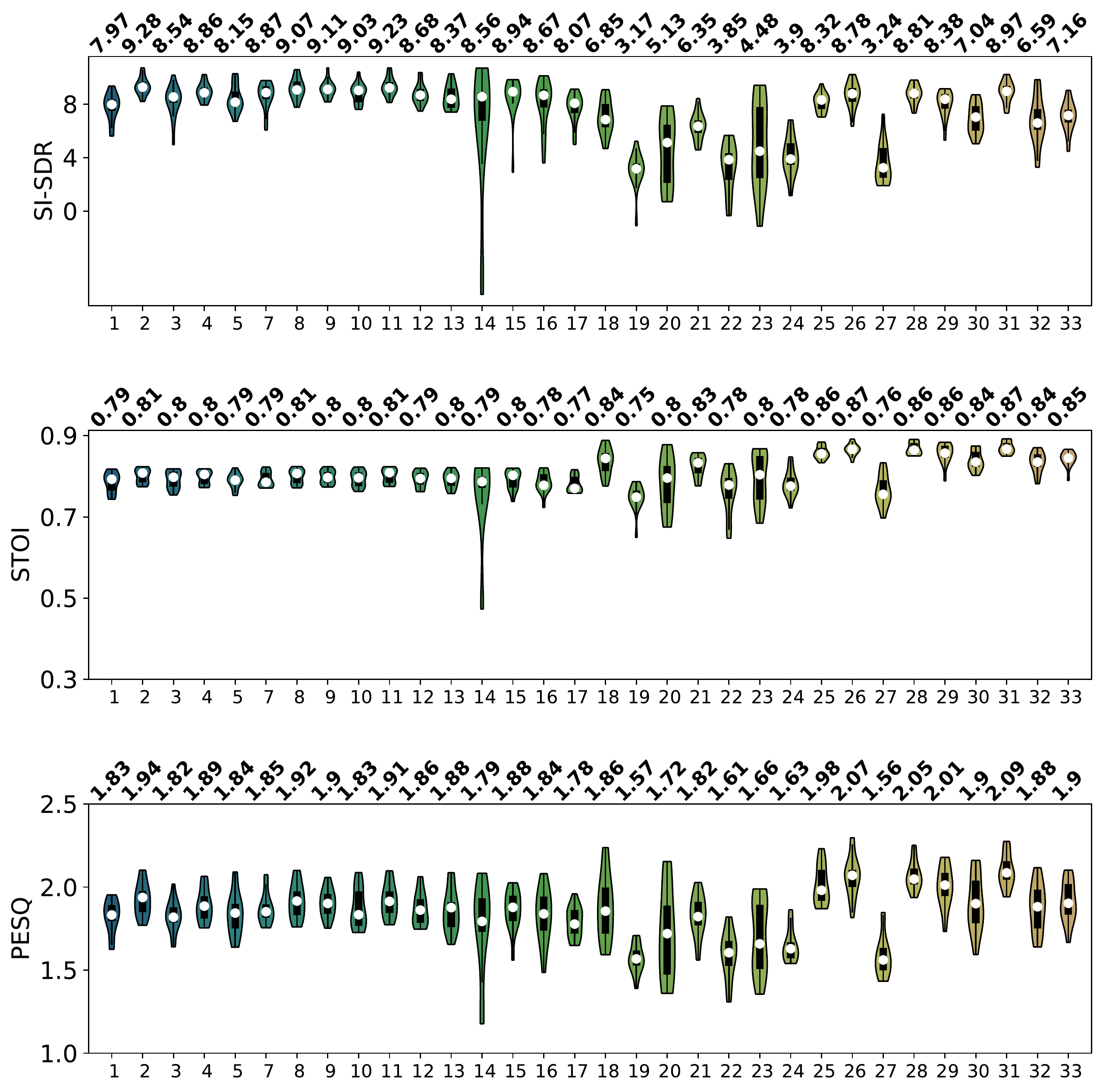}
    \vspace{-0.1cm}
}
  \subfigure[Proposed BASEN]{
  \centering
    \includegraphics[width=0.45\textwidth,height=0.4\textwidth]{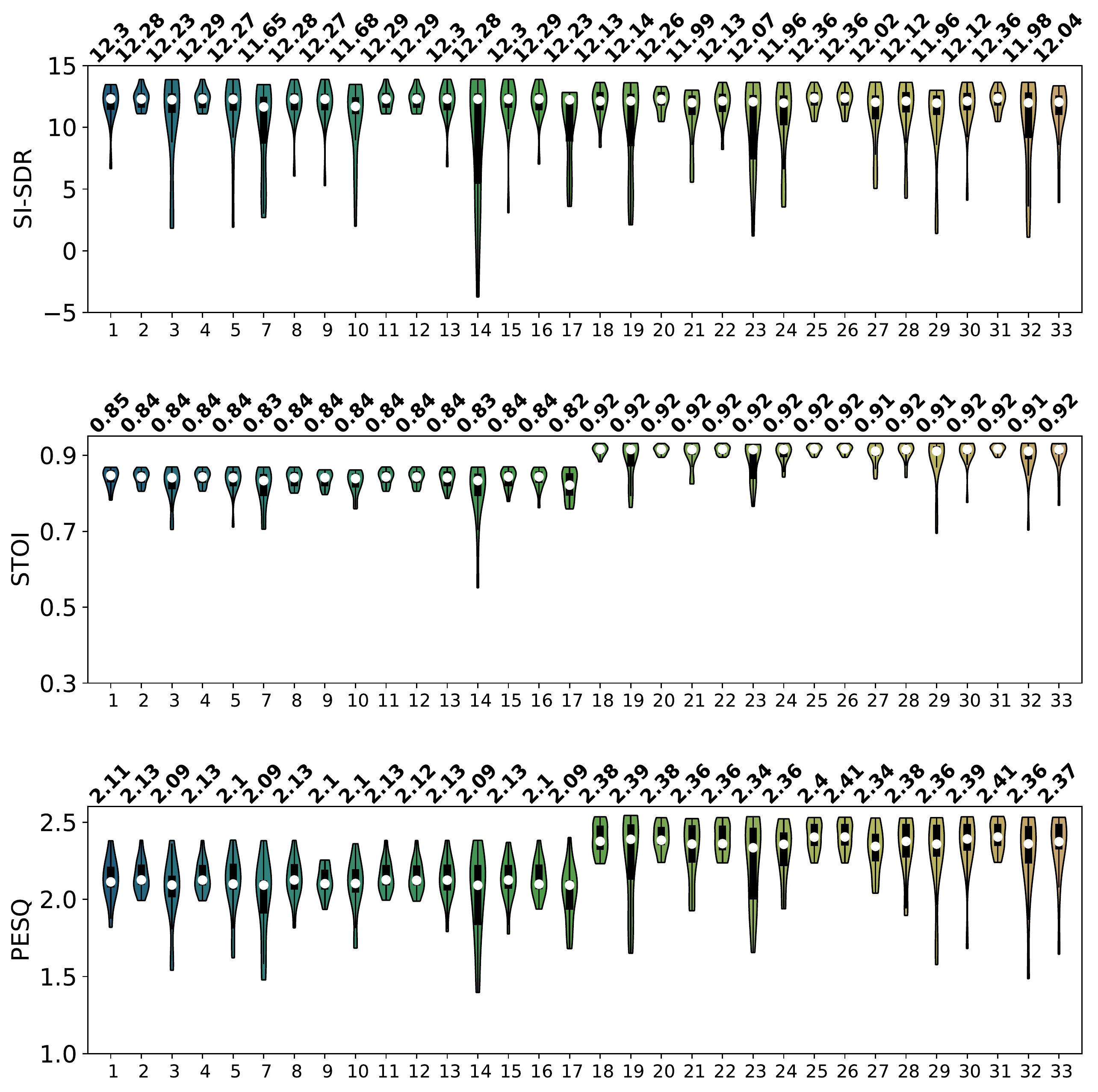}
    \vspace{-0.1cm}
}
    \vspace{-0.45cm}
  \caption{The comparison with the state-of-the-art UBESD method, where the median values are shown at the top of each sub-figure. The white dot in each plot shows the median. The black bar in the center of the violins shows the interquartile range (IQR). The thin black lines stretched from the bar show {\it first quartile $-1.5\times IQR$} and {\it third quartile} $+1.5\times IQR$, respectively.}
  \label{Figure3}
      \vspace{-0.5cm}
\end{figure*}

    \vspace{-0.1cm}
\section{Experimental setup}
    \vspace{-0.1cm}
{\it \textbf{Dataset:}}
The dataset used in this work was obtained from  \cite{broderick2018electrophysiological}. For this dataset, all procedures were performed in accordance with the Declaration of Helsinki and were approved by the Ethics Committees of Trinity College Dublin. All subjects were native English normal-hearing speakers without any history of neurological diseases. A total of 33 subjects (28 males and 5 females) with a mean age of 27.3$\pm$3.2 years participated in the experiments, but the data from subject \#6 was not included in the analysis due to the poor quality.

The subjects undertook 30 trials and each contains 60 seconds. During each trial, they were presented with two stories, one to the left ear and the other to the right ear. Each story was read by a different male speaker. Subjects were divided into two groups, and each group was instructed to pay attention to either the left (17) or the right ear (16 + 1 excluded subject) with each group instructed to attend to the story in either the left or right ear throughout the entire 30 trials. After each trial, subjects were required to answer multiple choice questions on both stories to test their attention. The storyline was continuous, that is, for each trial the story began where the last trial ended.

The stimuli amplitudes were normalized to have the same root mean square (RMS) level and silent gaps were cut short to a maximum of 0.5 seconds. Stimuli were presented using Sennheiser HD650 headphones and a presentation software from Neurobehavioral Systems with a sampling frequency of 44.1 kHz. Subjects were required to maintain a visual fixation on a cross hair centered on the screen and to minimize eye blinking and other motor activities.
EEG data were recorded using a 128-channel (plus two mastoids) EEG cap at a rate of 512 Hz using a BioSemi ActiveTwo system. The EEG recordings were further downsampled to 128 Hz. More details about the experimental setting can be found in \cite{broderick2018electrophysiological}.

{\it \textbf{Pre-processing:}}
Our experimental setup is kept the same as that in~\cite{hosseini_end--end_2022}. In order to reduce the computational complexity, we downsampled the sound data to a sampling rate of 14.7 kHz. The two stimuli were normalized to have the same RMS level and then equally added to synthesize the noisy mixture at a fixed SNR of 0 dB.  That is, the target speaker and the interfering source have the same power, and the presence of additional background noise or reverberation was not taken into account.  We divided the data into the following three groups: randomly choosing 5 trials from all subjects as the test data and 2 trials as the validation data, and all the remaining trials were used as the training data. For the training and validation sets, each trial was cut into 2-second segments. For the testing set, each 60-second trial was cut into 20-second segments.

The EEG data were first pre-processed using a bandpass filter with the pass frequency ranging from  0.1 Hz to 45 Hz, such that only related frequency bands are preserved and the electrical noise (50 or 60 Hz) and very low-frequency noise originating from the drift in the recording environment can be removed. To identify channels with excessive noise, the standard deviation (SD) of each channel was compared to the SD of the surrounding channels and each channel was visually inspected. Channels with excessive noise were recalculated by spline interpolation of the surrounding channels. The EEGs were re-referenced to the average of the mastoid channels to avoid introducing noise from the reference site. To remove artifacts caused by eye blinking and other muscle movements, we performed independent component analysis (ICA), which was done in EEGLAB~\cite{delorme2004eeglab}. For each subject, the trial that contains too much noise was excluded from the experiments.

In principle, EEG signals are noisy mixtures of potential existing sources that are not necessarily related to the target stimuli. It was shown in~\cite{hosseini_end--end_2022} that using the underlying neural activity more related to the speech stimuli for the separation network is more beneficial for the performance than directly using the EEG signals.
We therefore further processed the EEG data using a frequency-band coupling model to extract the audio related information in the EEG data, which can be represented by the cortical multiunit neural activity (MUA) from EEG signals. In \cite{moinnereau_frequency-band_2020,whittingstall_frequency-band_2009}, MUA was shown to be  effective for estimating the neural activity in the visual and auditory systems,  given by
\begin{equation}
    N(t)=a_{\gamma}\times P_{\gamma}(t)+a_{\delta}\times \angle \delta(t)
\end{equation}
where $P_{\gamma}(t)$ and $\angle \delta(t)$ are the amplitude of the gamma band and the phase of the delta band, respectively, and in experiments both $a_{\gamma}$ and $a_{\delta}$  are set to be 0.5 accordingly.

In addition, we use three objective metrics to evaluate the SE performance, including SI-SDR \cite{le2019sdr} in dB, perceptual evaluation of speech quality (PESQ) \cite{rix2001perceptual} and short-time objective intelligibility (STOI)~\cite{taal2010short}.
For training, the Adam optimizer is set with a momentum of $\beta_{1}$ = 0.9 and a denominator momentum of $\beta_{2}$ = 0.999. We use the linear warmup following the cosine annealing learning rate schedule with a maximum learning rate of $2\times 10^{-4}$ and a warmup ratio of $5\%$. The model is trained with around  60 epochs and a batch size of 8.

\begin{figure}[!t]
\centering
  \subfigure[Ablation study]{ 
   \centering
    \includegraphics[width=0.223\textwidth]{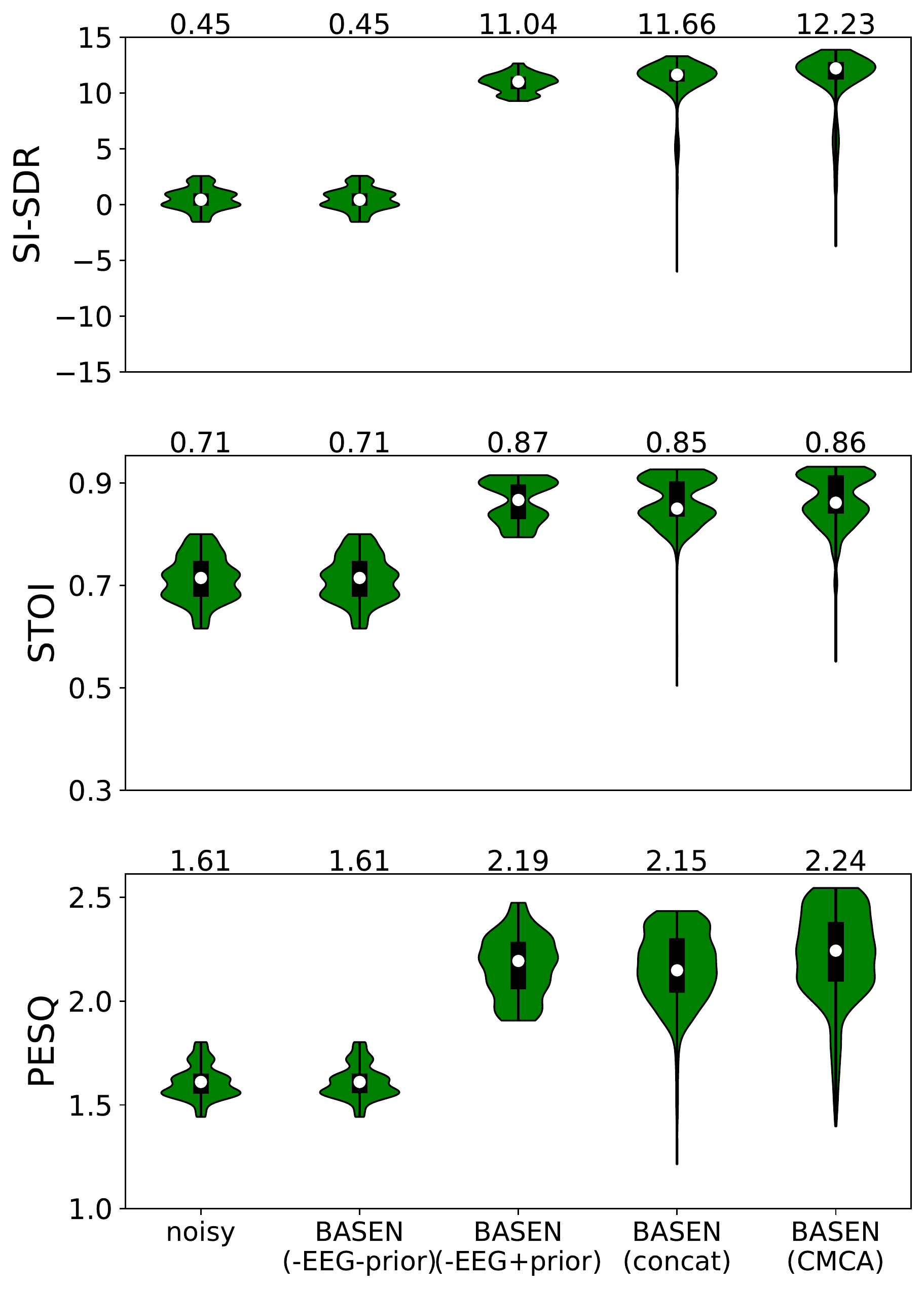}
    \vspace{-0.1cm}
  }
  \subfigure[Impact of CMCA layers]{
     \centering
    \includegraphics[width=0.223\textwidth]{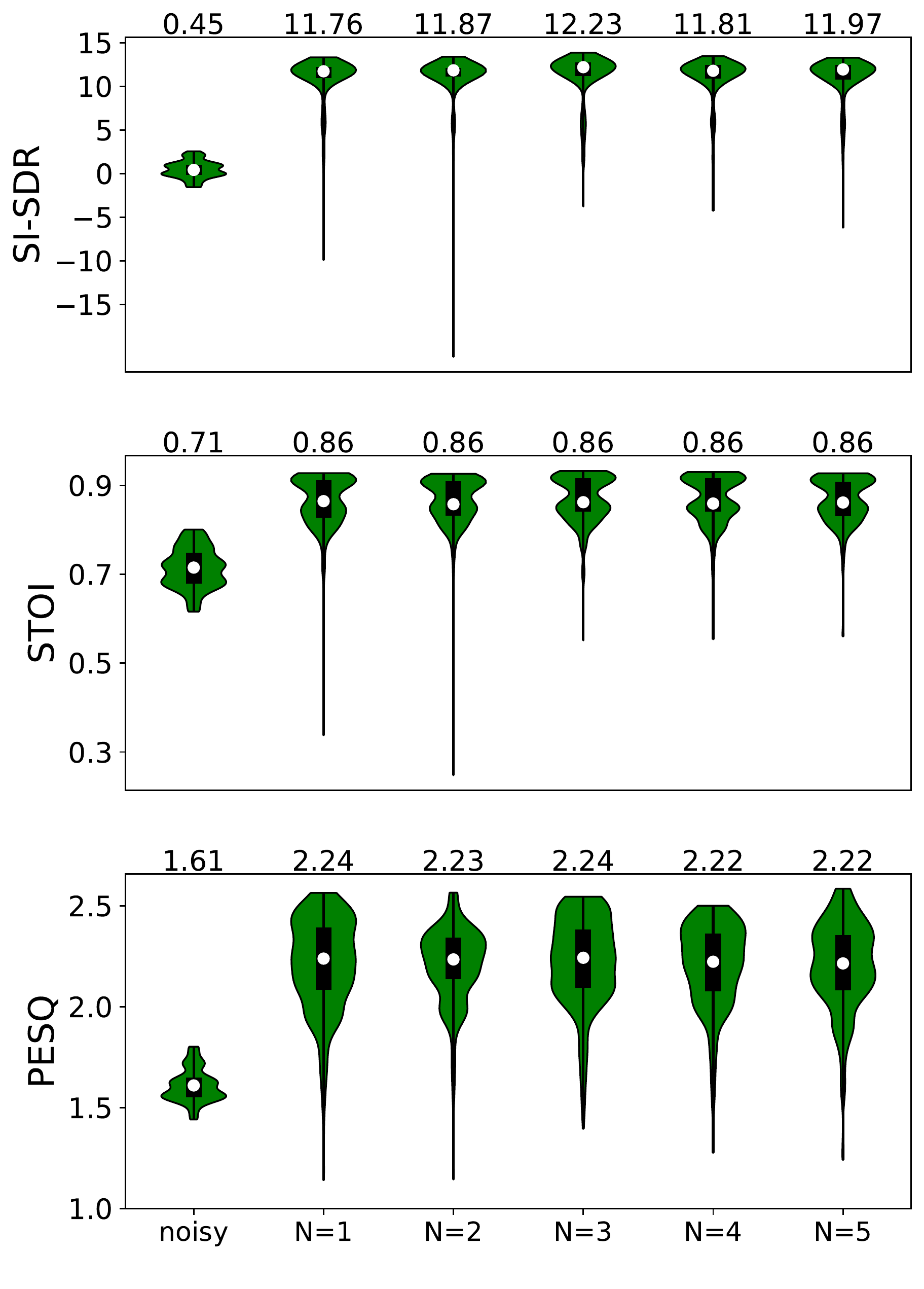}
    \vspace{-0.1cm}
   }
       \vspace{-0.5cm}
  \caption{The SE performance  for unknown attended speaker denoising in terms of SI-SDR , PESQ and STOI.}
  \label{fig:self-comparison}
      \vspace{-0.5cm}
\end{figure}

    \vspace{-0.1cm}
\section{Experimental results}
    \vspace{-0.1cm}
In this section, we first conduct several experiments to show the effectiveness of the designed modules within the proposed BASEN model. In Figure~\ref{fig:self-comparison}(a), we show the performance of BASEN using the audio-only signal, simple concatenation and CMCA, respectively.  It is clear that without prior information on the attended speaker, the typical  Conv-TasNet (i.e., BASEN-EEG-prior) cannot improve the speech quality, since both speakers produce speech signals and no ambient noises are present  in the considered  setting. Given the attended speaker, Conv-TasNet becomes equivalent to the proposed BASEN without EEG trials but with prior information, which can largely improve the performance.  This shows that the efficacy of Conv-TasNet~\cite{luo_conv-tasnet_2019} on SE is heavily dependent on the prior attended speaker information.  Including the EEG branch and simply concatenating the audio and EEG embeddings can achieve a comparable performance as the ideal Conv-TasNet. Applying the proposed CMCA module for feature fusion further improves the performance, particularly the PESQ score.

In order to find out the most appropriate number of cross-attention layers in the CMCA module, we  evaluate the impact of the layer number on the performance in Figure~\ref{fig:self-comparison}(b), where $N$ changes from 1 to 5.  It is clear the choice of $N$ = 3 returns the best SE performance in all metrics, which will be used in the sequel. Note that including more layers will also increase the parameter amount of the overall BASEN model, which increases from 0.57M for $N$ = 1 to 0.64M for $N$ = 3 for example.

Then, we conduct a comparison  of the proposed BASEN with the best published method on the same dataset, i.e., UBESD in~\cite{hosseini_end--end_2022}. The obtained SI-SDR, PESQ and STOI in terms of the testing subjects are summarized in Figure~\ref{Figure3}. We can clearly see that  the proposed BASEN outperforms the state-of-the-art UBESD in all  metrics and over all subjects. More importantly, the performance of BASEN is more subject invariant, as that of UBESD varies more across subjects, meaning that BASEN is more robust against listening dynamics. This also shows that the proposed CMCA module is more effective for feature fusion than the FiLM strategy in~\cite{hosseini_end--end_2022}. Note that  UBESD  also has a larger parameter amount, which is around 1.84M, due to the fact that the adopted FiLM has four convolutions and each has more channels. 
    \vspace{-0.1cm}
\section{Conclusion}
    \vspace{-0.1cm}
In this paper, we proposed the BASEN approach for time-domain SE using the EEG signal of the listener as an additional modality, where the CMCA module was designed to deeply fuse the audio and EEG embeddings. It was shown that the EEG trials are helpful for extracting the attended speaker, and the proposed BASEN method can improve the speech quality without any prior information on the target speaker. This would be a useful candidate for realistic applications where no prior information on the attended speaker is given. Due to the page limitation, more results will be presented in a future journal version.

\bibliographystyle{IEEEtran}
\bibliography{refs}

\begin{thebibliography}{10}
\providecommand{\url}[1]{#1}
\csname url@samestyle\endcsname
\providecommand{\newblock}{\relax}
\providecommand{\bibinfo}[2]{#2}
\providecommand{\BIBentrySTDinterwordspacing}{\spaceskip=0pt\relax}
\providecommand{\BIBentryALTinterwordstretchfactor}{4}
\providecommand{\BIBentryALTinterwordspacing}{\spaceskip=\fontdimen2\font plus
\BIBentryALTinterwordstretchfactor\fontdimen3\font minus
  \fontdimen4\font\relax}
\providecommand{\BIBforeignlanguage}[2]{{%
\expandafter\ifx\csname l@#1\endcsname\relax
\typeout{** WARNING: IEEEtran.bst: No hyphenation pattern has been}%
\typeout{** loaded for the language `#1'. Using the pattern for}%
\typeout{** the default language instead.}%
\else
\language=\csname l@#1\endcsname
\fi
#2}}
\providecommand{\BIBdecl}{\relax}
\BIBdecl

\bibitem{Jinyu-et-al:scheme}
J.~Li, L.~Deng, R.~Haeb-Umbach, and Y.~Gong, \emph{Robust automatic speech
  recognition: a bridge to practical applications}.\hskip 1em plus 0.5em minus
  0.4em\relax Oxford: Academic Press, 2016.

\bibitem{zhao2018deep}
Y.~Zhao, D.~Wang, E.~M. Johnson, and E.~W. Healy, ``A deep learning based
  segregation algorithm to increase speech intelligibility for hearing-impaired
  listeners in reverberant-noisy conditions,'' \emph{J. of the Acoust. Soc.
  Am.}, vol. 144, no.~3, pp. 1627--1637, 2018.

\bibitem{krishna_advancing_2020}
G.~Krishna, C.~Tran, Y.~Han, M.~Carnahan, and A.~H. Tewfik, ``Speech synthesis
  using {EEG},'' in \emph{ICASSP}, 2020, pp. 1235--1238.

\bibitem{benesty2008microphone}
J.~Benesty, J.~Chen, and Y.~Huang, \emph{Microphone array signal
  processing}.\hskip 1em plus 0.5em minus 0.4em\relax Springer Science \&
  Business Media, 2008, vol.~1.

\bibitem{loizou2007speech}
P.~C. Loizou, \emph{Speech enhancement: theory and practice}, 2007.

\bibitem{wang2014training}
Y.~Wang, A.~Narayanan, and D.~Wang, ``On training targets for supervised speech
  separation,'' \emph{IEEE/ACM Trans. Audio, Speech, Lang. Process.}, vol.~22,
  no.~12, pp. 1849--1858, 2014.

\bibitem{erdogan2015phase}
H.~Erdogan, J.~R. Hershey, S.~Watanabe, and J.~Le~Roux, ``Phase-sensitive and
  recognition-boosted speech separation using deep recurrent neural networks,''
  in \emph{ICASSP}, 2015, pp. 708--712.

\bibitem{xu2014regression}
Y.~Xu, J.~Du, L.-R. Dai, and C.-H. Lee, ``A regression approach to speech
  enhancement based on deep neural networks,'' \emph{IEEE/ACM Trans. Audio,
  Speech, Lang. Proce.}, vol.~23, no.~1, pp. 7--19, 2014.

\bibitem{tan_learning_2020}
K.~Tan and D.~Wang, ``Learning complex spectral mapping with gated
  convolutional recurrent networks for monaural speech enhancement,''
  \emph{IEEE/ACM Trans. Audio, Speech, Lang. Process.}, vol.~28, pp. 380--390,
  2019.

\bibitem{williamson_time-frequency_2017}
D.~S. Williamson and D.~Wang, ``Time-frequency masking in the complex domain
  for speech dereverberation and denoising,'' \emph{IEEE/ACM Trans. Audio,
  Speech, Lang. Process.}, vol.~25, no.~7, pp. 1492--1501, 2017.

\bibitem{wang2021compensation}
Z.-Q. Wang, G.~Wichern, and J.~Le~Roux, ``On the compensation between magnitude
  and phase in speech separation,'' \emph{IEEE Signal Process. Letters},
  vol.~28, pp. 2018--2022, 2021.

\bibitem{li2022taylor}
A.~Li, S.~You, G.~Yu, C.~Zheng, and X.~Li, ``Taylor, can you hear me now? a
  taylor-unfolding framework for monaural speech enhancement,'' in \emph{Proc.
  IJCAI}, 2022, pp. 4193--4200.

\bibitem{pandey_dense_2021}
A.~Pandey and D.~Wang, ``Dense cnn with self-attention for time-domain speech
  enhancement,'' \emph{IEEE/ACM Trans. Audio, Speech, Lang. Process.}, vol.~29,
  pp. 1270--1279, 2021.

\bibitem{luo_conv-tasnet_2019}
Y.~Luo and N.~Mesgarani, ``Conv-{TasNet}: {Surpassing} {Ideal}
  {Time}-{Frequency} {Magnitude} {Masking} for {Speech} {Separation},''
  \emph{IEEE/ACM Trans. Audio, Speech, Lang. Process.}, vol.~27, no.~8, pp.
  1256--1266, 2019.

\bibitem{luo2018tasnet}
------, ``Tasnet: time-domain audio separation network for real-time,
  single-channel speech separation,'' in \emph{ICASSP}, 2018, pp. 696--700.

\bibitem{pandey2019new}
A.~Pandey and D.~Wang, ``A new framework for {CNN}-based speech enhancement in
  the time domain,'' \emph{IEEE/ACM Trans. Audio, Speech, Lang. Process.},
  vol.~27, no.~7, pp. 1179--1188, 2019.

\bibitem{hu2021speech}
X.~Hu, K.~Li, W.~Zhang, Y.~Luo, J.~Lemercier, and T.~Gerkmann, ``Speech
  separation using an asynchronous fully recurrent convolutional neural
  network,'' \emph{NIPS}, vol.~34, pp. 22\,509--22\,522, 2021.

\bibitem{mesgarani_selective_2012}
N.~Mesgarani and E.~F. Chang, ``Selective cortical representation of attended
  speaker in multi-talker speech perception,'' \emph{Nature}, vol. 485, no.
  7397, p. 10.1038/nature11020, 2012.

\bibitem{osullivan_attentional_2015}
J.~A. O'Sullivan, A.~J. Power, N.~Mesgarani, and et~al., ``Attentional
  selection in a cocktail party environment can be decoded from single-trial
  {EEG},'' \emph{Cerebral Cortex}, vol.~25, no.~7, pp. 1697--1706, 2015.

\bibitem{aroudi_auditory_2016}
A.~Aroudi, B.~Mirkovic, M.~De~Vos, and S.~Doclo, ``Auditory attention decoding
  with {EEG} recordings using noisy acoustic reference signals,'' in
  \emph{ICASSP}, 2016, pp. 694--698.

\bibitem{osullivan_neural_2017}
J.~O'Sullivan, Z.~Chen, J.~Herrero, G.~M. McKhann, S.~A. Sheth, A.~D. Mehta,
  and N.~Mesgarani, ``Neural decoding of attentional selection in multi-speaker
  environments without access to clean sources,'' \emph{J. Neural Engineering},
  vol.~14, no.~5, p. 056001, 2017.

\bibitem{aroudi2021closed}
A.~Aroudi, E.~Fischer, M.~Serman, H.~Puder, and S.~Doclo, ``Closed-loop
  cognitive-driven gain control of competing sounds using auditory attention
  decoding,'' \emph{Algorithms}, vol.~14, no.~10, p. 287, 2021.

\bibitem{geirnaert_electroencephalography-based_2021}
S.~Geirnaert, S.~Vandecappelle, E.~Alickovic, and et~al.,
  ``Electroencephalography-based auditory attention decoding: Toward
  neurosteered hearing devices,'' \emph{IEEE Signal Process. Mag.}, vol.~38,
  no.~4, pp. 89--102, 2021.

\bibitem{han2019speaker}
C.~Han, J.~O'™Sullivan, Y.~Luo, J.~Herrero, A.~D. Mehta, and N.~Mesgarani,
  ``Speaker-independent auditory attention decoding without access to clean
  speech sources,'' \emph{Science advances}, vol.~5, no.~5, p. eaav6134, 2019.

\bibitem{wang2018voicefilter}
Q.~Wang, H.~Muckenhirn, K.~Wilson, Z.~Sridhar, P.and~Wu, J.~Hershey, R.~A.
  Saurous, R.~J. Weiss, Y.~Jia, and I.~L. Moreno, ``Voicefilter: Targeted voice
  separation by speaker-conditioned spectrogram masking,'' \emph{arXiv preprint
  arXiv:1810.04826}, 2018.

\bibitem{ochiai2019multimodal}
T.~Ochiai, M.~Delcroix, K.~Kinoshita, A.~Ogawa, and T.~Nakatani, ``Multimodal
  speakerbeam: Single channel target speech extraction with audio-visual
  speaker clues.'' in \emph{ISCA Interspeech}, 2019, pp. 2718--2722.

\bibitem{li2020listen}
C.~Li and Y.~Qian, ``Listen, watch and understand at the cocktail party:
  Audio-visual-contextual speech separation.'' in \emph{ISCA Interspeech},
  2020, pp. 1426--1430.

\bibitem{michelsanti2021overview}
D.~Michelsanti, Z.-H. Tan, S.-X. Zhang, Y.~Xu, M.~Yu, D.~Yu, and J.~Jensen,
  ``An overview of deep-learning-based audio-visual speech enhancement and
  separation,'' \emph{IEEE/ACM Trans. Audio, Speech, Lang. Process.}, vol.~29,
  pp. 1368--1396, 2021.

\bibitem{ceolini_brain-informed_2020}
E.~Ceolini, J.~Hjortkj{\ae}r, D.~D. Wong, J.~O'Sullivan, V.~S. Raghavan,
  J.~Herrero, A.~D. Mehta, S.-C. Liu, and N.~Mesgarani, ``Brain-informed speech
  separation ({BISS}) for enhancement of target speaker in multitalker speech
  perception,'' \emph{NeuroImage}, vol. 223, p. 117282, 2020.

\bibitem{hosseini_end--end_2022}
M.~Hosseini, L.~Celotti, and E.~Plourde, ``End-to-{End} {Brain}-{Driven}
  {Speech} {Enhancement} in {Multi}-{Talker} {Conditions},'' \emph{IEEE/ACM
  Trans. Audio, Speech, and Lang. Process.}, vol.~30, pp. 1718--1733, 2022.

\bibitem{vaswani2017attention}
A.~Vaswani, N.~Shazeer, N.~Parmar, J.~Uszkoreit, L.~Jones, A.~N. Gomez,
  {\L}.~Kaiser, and I.~Polosukhin, ``Attention is all you need,'' \emph{NIPS},
  vol.~30, 2017.

\bibitem{le2019sdr}
J.~Le~Roux, S.~Wisdom, H.~Erdogan, and J.~R. Hershey, ``Sdr--half-baked or well
  done?'' in \emph{ICASSP}, 2019, pp. 626--630.

\bibitem{kolbaek2020loss}
M.~Kolb{\ae}k, Z.-H. Tan, S.~H. Jensen, and J.~Jensen, ``On loss functions for
  supervised monaural time-domain speech enhancement,'' \emph{IEEE/ACM Trans.
  Audio, Speech, Lang. Process.}, vol.~28, pp. 825--838, 2020.

\bibitem{broderick2018electrophysiological}
M.~P. Broderick, A.~J. Anderson, G.~M. Di~Liberto, M.~J. Crosse, and E.~C.
  Lalor, ``Electrophysiological correlates of semantic dissimilarity reflect
  the comprehension of natural, narrative speech,'' \emph{Current Biology},
  vol.~28, no.~5, pp. 803--809, 2018.

\bibitem{delorme2004eeglab}
A.~Delorme and S.~Makeig, ``{EEGLAB}: an open source toolbox for analysis of
  single-trial {EEG} dynamics including independent component analysis,''
  \emph{J. Neuroscience Methods}, vol. 134, no.~1, pp. 9--21, 2004.

\bibitem{moinnereau_frequency-band_2020}
M.~Moinnereau, J.~Rouat, K.~Whittingstall, and E.~Plourde, ``A frequency-band
  coupling model of {EEG} signals can capture features from an input audio
  stimulus,'' \emph{Hearing Research}, vol. 393, p. 107994, 2020.

\bibitem{whittingstall_frequency-band_2009}
K.~Whittingstall and N.~K. Logothetis, ``Frequency-band coupling in surface
  {EEG} reflects spiking activity in monkey visual cortex,'' \emph{Neuron},
  vol.~64, no.~2, pp. 281--289, 2009.

\bibitem{rix2001perceptual}
A.~W. Rix, J.~G. Beerends, M.~P. Hollier, and A.~P. Hekstra, ``Perceptual
  evaluation of speech quality ({PESQ})-a new method for speech quality
  assessment of telephone networks and codecs,'' in \emph{ICASSP}, vol.~2,
  2001, pp. 749--752.

\bibitem{taal2010short}
C.~H. Taal, R.~C. Hendriks, R.~Heusdens, and J.~Jensen, ``A short-time
  objective intelligibility measure for time-frequency weighted noisy speech,''
  in \emph{ICASSP}, 2010, pp. 4214--4217.

\end{thebibliography}

\end{document}